\documentclass[twocolumn,prb,aps]{revtex4}
\usepackage{epsfig}
\usepackage{amsmath}
\usepackage{amssymb}

\begin{document}
\textheight 226.5mm

\title{Influence of Mo on the Fe:Mo:C nano-catalyst thermodynamics for single-walled carbon nanotube growth} 
\author{Stefano Curtarolo$^{1,*}$, Neha Awasthi$^1$, Wahyu Setyawan$^1$, Aiqin Jiang$^1$, \\Kim Bolton$^2$, Toshio Tokune$^3$, Avetik R. Harutyunyan$^3$} 
\affiliation{
  $^1${\small Department of Mechanical Engineering and Materials Science Duke University Durham, NC 27708}\\
  $^2${\small University College of Boraas SE-501 90 Boraas and Physics Department G\"oteborg University SE-412 96 G\"oteborg, Sweden}\\
  $^3${\small Honda Research Institute USA Inc. 1381 Kinnear Road Columbus, OH 43212}\\
  $^*$corresponding author: stefano@duke.edu }

\date{\today}


\begin{abstract}
We explore the role of Mo in Fe:Mo nanocatalyst thermodynamics 
for low-temperature chemical vapor deposition growth of single walled carbon nanotubes (SWCNTs).
By using the size-pressure approximation and {\it ab initio} modeling,
we prove that for both Fe-rich ($\sim$ 80\% Fe or more) and Mo-rich ($\sim$ 50\% Mo or more) Fe:Mo clusters, 
the presence of carbon in the cluster causes nucleation of Mo$_2$C.  This 
enhances the activity of the particle since it
releases Fe, which is initially bound in a stable Fe:Mo phase, so that it can catalyze SWCNT growth.
Furthermore, the presence of small concentrations of Mo 
reduce the lower size limit of low-temperature steady-state growth from $\sim$0.58$nm for pure Fe particles to \sim0.52$nm. 
Our {\it ab initio}-thermodynamic modeling explains experimental results
and establishes a new direction to search for better catalysts.
\end{abstract}

\maketitle


Critical factors for the efficient growth of single walled carbon nanotubes
(SWCNTs) via catalytic chemical vapor deposition (CCVD) \cite{exp1,exp2,exp2c}
are the compositions of the interacting species (feedstock, catalyst, support 
\cite{Mosaia_JPCM_2003,Kakehi_CPL_2006,expDai,expKitiyanan,expCheung,expZhu,expHarutyunyan}), the preparation of the 
catalysts, and the synthesis conditions \cite{expDupuis,expKanzow,expCantoro,expKlinke,expLin,expCassel,expHata,expZhang}.
Efficient catalysts must have long active lifetimes 
(with respect to feedstock dissociation and nanotube growth),
high selectivity and be less prone to contamination 
\cite{Mora_CARBON_2007,Avetik_APL_2007,Jiang_PRB_2007,Curtarolo_SOLUBILITY1,Curtarolo_SOLUBILITY2}.
Common factors that lead to reduction in catalytic activity are 
deactivation (i.e. chemical poisoning or coating with carbon),
thermal sintering (e.g. caused by highly exothermic reactions on the clusters surface \cite{expCantoro,exp4,expKlinke} 
with insufficient heat transfer \cite{Pernicone,WLWang})
and solid-state reactions (nucleation of inactive phases in the cluster \cite{Pernicone,Curtarolo_SOLUBILITY1,Curtarolo_SOLUBILITY2}).

Metal alloy catalysts, such as Fe:Co, Co:Mo and Fe:Mo, improve the growth of CNTs 
\cite{Mosaia_JPCM_2003,Avetik_APL_2007,Shah1,Shah2,expHarutyunyan,Flahaut_CPL_1999,Ago,Tang}, 
because the presence of more than one metal species can significantly enhance the activity of a catalyst 
\cite{Avetik_APL_2007,Resasco_CoMo1,Resasco_CoMo2,Lamouroux_FeMo}, and can prevent catalyst particle aggregation 
\cite{Avetik_APL_2007,Tang,Resasco_CoMo1,Resasco_CoMo2}.
In the case of Fe:Mo nanoparticles supported on Al$_2$O$_3$ substrates, 
the enhanced catalyst activity has been shown to be larger than the 
linear combination of the individual Fe/Al$_2$O$_3$ and Mo/Al$_2$O$_3$ activities \cite{Avetik_APL_2007,Shah1,Shah2}.
This is explained in terms of substantial inter-metallic interaction between 
Mo, Fe and C \cite{Avetik_APL_2007,Massalski,PAULING} which is congruent with 
previously observed solid-state reactions between these elements. 
In fact, the addition of Mo in mechanical alloying of powder Fe and C mixtures \cite{Omuro} promotes solid state 
reactions even at low Mo concentrations by forming ternary phases, such as the (Fe,Mo)$_{23}$C$_6$ type carbides \cite{Omuro}.

The way in which carbon interacts with transition metals depends on the metal species. 
Fe and Co belong to the ``carbon dissolution-precipitation mechanism'' group, where relatively large fractions of 
carbon dissolve into the cluster before stable carbides are formed, 
while Mo belongs to the ``carbide formation-decomposition'' group where carbide 
formation occurs rapidly at low carbon concentrations \cite{Oya}.
These mechanisms are governed by the interplay between solubility of C in the
metal matrix (interstitial and/or substitutional defects) and the ease of metal carbide formation. 
Although relating C solubility and catalytic ability of metal catalysts is not simple 
(factors like temperature, diffusion, kinetics and particle size 
have to be considered \cite{Curtarolo_SOLUBILITY1,Curtarolo_SOLUBILITY2}), 
metals which form carbides can be efficient
catalysts as long as they offer sufficient carbon solubility, 
even if the formation of the carbides interferes with the SWCNT growth, as is the 
case of Fe \cite{Curtarolo_SOLUBILITY1,Curtarolo_SOLUBILITY2}.
Co can dissolve a relatively large fraction of carbon C and does not form stable ordered carbides, whereas
Mo has limited C solubility and forms several stable carbides (even at low temperature).
Similarly to Co, Fe  can dissolve a relatively large fraction of C 
(at least in the $\gamma$ phase), but concomitantly it forms carbides at temperatures 
relevant to SWCNT growth (Fe$_3$C) \cite{Massalski,PAULING}.
Nucleation of Fe$_3$C
and subsequent loss of C solubility\cite{Avetik_APL_2007} can be used to identify the thermodynamic lower limit 
for the CVD growth of very thin SWCNTs from Fe nanoparticles \cite{Curtarolo_SOLUBILITY1,Curtarolo_SOLUBILITY2}.
Thus, for efficient CVD growth from Fe, Mo, Fe:Mo and Co:Mo clusters, 
the particles must be small (dispersed) and metallic (preserved from carbidization).
In fact, active catalyst species in Co:Mo catalysis are also small metallic
species \cite{Resasco_CoMo1,Resasco_CoMo2} with 
Mo added to limit particle aggregation by
forming molybdate species which are later reduced to give small
metallic Co:Mo nanoparticles. It has been shown that 
disruption of the Co-Mo interactions leads to a lower SWCNT 
growth efficiency \cite{Resasco_CoMo3}.

The use of alloys instead of pure metals provides additional degrees of freedom, 
such as fractions of the metal species, which
has {\it chemical} and {\it thermodynamic} advantages. 
The chemical advantages arise since 
the fractions of constituent species can be tailored to enhance catalytic performance 
that can be measured by the
yield and quality of nanotubes \cite{Resasco_CoMo1,Resasco_CoMo2,Lamouroux_FeMo}.
It has been found that low Fe:Mo ratios are favorable for growing SWCNTs 
(on Al$_2$O$_3$ substrates) since the presence, after activation,
of the phase Fe$_2$(MoO$_4$)$_3$ can lead to the formation 
of small metallic clusters \cite{Lamouroux_FeMo} 
(the best compromise between catalytic activity and SWCNT selectivity 
was found to be Fe$_{3.5}$Mo$_{11.5}$, Ref. \cite{Lamouroux_FeMo}).
High fractions of Fe in Fe:Mo lead to formation of larger particles during
the reduction step which are inactive for SWCNT growth, unless 
precautions are taken to avoid excessive sintering \cite{Avetik_APL_2007} 
(no formation of molybdate species
has been reported when using CH$_4$ feedstock with Fe:Mo catalysts \cite{Avetik_APL_2007} 
and hence the chemical role of 
Mo in Fe:Mo is different from that in Co:Mo).
The thermodynamic advantages are revealed when  
considering the vapor-liquid-solid model (VLS), which is 
the most probable mechanism for CNT growth \cite{VLS1,expKanzow,Avetik_APL_2007}.
The metallic nanoparticles are very efficient catalysts when they are 
in the liquid or viscous states \cite{Avetik_APL_2007} ,
probably because one has considerable carbon bulk-diffusion in this phase 
(compared to surface or sub-surface diffusion).
Generally, unless stable intermetallic compounds form,
alloying metals reduce the melting point 
below those of the constituents \cite{Massalski,PAULING}.
This happens, for instance, with the addition of small fractions of Mo to Fe.
Hence, to enhance the yield and quality of nanotubes,
one can tailor the composition of the catalyst particle to move its {\it liquidus} line 
below the synthesis temperature \cite{Avetik_APL_2007}.
However, identifying the perfect alloy composition is non trivial. 
In fact, the presence of more than two metallic species allows for the
possibility of different carbon pollution mechanisms by allowing 
thermodynamic promotion of ternary carbides.
So far, due to the very complex interplay between competing phases at the nano-scale (even for pure
Fe particles \cite{Curtarolo_SOLUBILITY1,Curtarolo_SOLUBILITY2}), the search for the proper catalyst composition has been 
empirical \cite{Avetik_APL_2007,Resasco_CoMo1,Resasco_CoMo2,Resasco_CoMo3,Lamouroux_FeMo}.

In the present manuscript we address the interaction between C and Mo:Fe nanoparticles. 
We use thermodynamic and quantum mechanical results to discuss 
the complexity of Fe- ($\sim$80\% Fe or more) and Mo-rich ($\sim$50\% Mo or more) Fe:Mo catalysts, previously 
addressed phenomenologically.
The results are useful for the development of efficient catalysts for nanotube
and graphene growth.

{\bf Methods.}
Investigating the behavior of C in Fe:Mo nanoparticles
requires an understanding of the interplay of the various phases of the Fe-Mo-C system at the nano-scale.
Determining the thermodynamic stability of different phases in nanoparticles of different sizes
with {\it ab initio} calculations is complicated and computationally expensive. 
In Ref. \cite{Curtarolo_SOLUBILITY1,Curtarolo_SOLUBILITY2} we have developed a simple model, called the
``size-pressure approximation'', which allows one to estimate the phase diagram at the nanoscale starting from bulk calculations 
under pressure.

{\it The size-pressure approximation.}
Surface curvature and superficial dangling bonds on nanoparticles
are responsible for internal stress fields which modify the atomic bond lengths. 
For spherical clusters, the phenomenon can be modeled with the Young-Laplace equation
$\Delta P=2\gamma/R $ where the parameter $\gamma$ (surface tension for liquid particles)
can be calculated with {\it ab initio} methods.
As a first approximation, where all surface effects that are not included in the curvature are neglected, 
the study of phase diagrams for spherical particles can be mapped onto the study 
of phase diagrams for bulk systems under the same pressure that is produced 
by the curvature. It is important to mention that $\gamma$ is not a real surface 
tension but an {\it ab initio} fitting parameter describing size-induced stress in nanoparticles.
In our case, it can be assumed that $\gamma$ is independent of the fraction of Mo 
(since the best Fe:Mo catalysts are Fe-rich) and C 
(since the amount of C in the catalytically active particle state is limited) \cite{expHarutyunyan,Avetik_APL_2007}.
Figure 2 of Ref. \cite{Curtarolo_SOLUBILITY1} shows the implementation of the 
``size-pressure approximation'' for Fe nanoparticles.
The idea is simple. Two interpolations are involved: 
``pressure versus distortion'' and ``distortion versus curvature'' 
are coupled to obtain a relation ``pressure versus curvature'' (where curvature is $1/R$ \cite{Curtarolo_SOLUBILITY1}).
In more details, the left hand side of Figure 2 of Ref. \cite{Curtarolo_SOLUBILITY1} 
shows the average distortion of the bond length inside the cluster $\Delta d_{nn}\equiv d_{nn}^{0}-d_{nn}$ 
for a variety of spherical bcc particles as a function of the inverse radius ($1/R$).
The right hand side shows the compression of the bond length as a
function of the hydrostatic pressure for the bulk system. 
Combining the two sets of data yields:
\begin{equation}
  P\cdot R=2.46 \,\,{\rm GPa} \cdot {\rm nm} \,\,\,\,(\gamma=1.23 {\rm J/m}^2). 
  \label{sizepressureequation}
\end{equation}
Equation (\ref{sizepressureequation}) is used to deduce the Fe-Mo-C phase diagram of nanoparticles of radius $R$ 
from {\it ab initio} calculations of the bulk material under pressure $P$.
The parameter $\gamma$ compares well with the
experimental values of the surface tension of bulk Fe at 
the melting point $\sim$ 1.85 J/m$^2$ \cite{gammaFEexp} and $\sim$1.90 \cite{gammaMOexp}.
For a detailed explanation of the ``size-pressure approximation'' see Ref. \cite{Curtarolo_SOLUBILITY1}.

The assumption of $\gamma$ being independent of the Mo concentration is justified 
at zero (and low) temperature because molybdenum tends to segregate inside the particle,
so that it can not affect the chemistry and the bonding states at the surface 
(see the section ``surface energies'' near the end of the manuscript).
At high temperature, Mo would eventually populate the surface and modify $\gamma$. 
To a first approximation, by linearly interpolating between the experimental values of the surface 
tension for Fe (1.85 J/m$^2$) and Mo (2.08 J/m$^2$) (i.e. Vegard's law), 
and assuming that the parameter $\gamma$ follows the same trend, 
we would obtain a $\sim$ 3\% increase for $\gamma$ at the optimal composition Fe$_4$Mo of the catalyst (see below).  
Thus, all the estimated radii and diameters deduced from equation \ref{sizepressureequation} 
might be overestimated of few percent at high temperature (energies and pressures are not affected by $\gamma$).
Considering the formidable computational needs required to obtain a concentration-dependent 
parameterization of $\gamma$, we believe that our calculated value is accurate enough 
to describe the general trends of size-induced thermodynamics and activity.

{\it Quantum mechanical calculations.} 
Simulations are performed with {\small VASP} \cite{kresse1993},
using projector augmented waves (PAW) \cite{bloechl994} and exchange-correlation functionals as 
parameterized by Perdew, Burke and Ernzerhof (PBE) \cite{PBE} for the generalized gradient approximation (GGA). 
Simulations are carried out with spin polarization, at zero temperature, and without zero-point motion.
All structures are fully relaxed.  Numerical convergence to within about 2 meV/atom is ensured by 
enforcing a high energy cut-off (500 eV) and dense {\bf k}-meshes.
The hydrostatic pressure estimated from the pressure-size model is implemented as
Pulay stress \cite{Pulay}.  
Ternary phase diagrams are calculated using bcc-Mo, bcc-Fe and SWCNTs as references
(pure-Fe phase is taken to be bcc because our simulations are aimed
at the low temperature regime of catalytic growth).
The reference SWCNTs have the same diameter of the particle 
to minimize the curvature-strain energy. In fact, CVD experiments of SWCNT growth from small 
($\sim$0.6-2.1 nm) particles indicate that the diameter of the nanotube is similar to the diameter of the catalyst
particle from which it grows. In some experiments where the growth
mechanism is thought to be root-growth, the ratio of the catalyst particle
diameter to SWCNT diameter is $\sim$1.0, whereas in experiments involving pre-made
floating catalyst particles this ratio is $\sim$1.6 \cite{Kim_Review_2008}.
Formation energies are calculated with respect to decomposition into the
nearby stable elements or phases, depending the position in the ternary phase diagram as described later.

{\it Competing phases.} 
Pure elements (bcc-Fe, bcc-Mo and SWCNT-C) are included in the calculation in their most stable 
low temperature configurations. Other phases are included if they are stable in the temperature range used in CVD
growth of SWCNTs or if they have been reported experimentally during or after the growth 
\cite{Massalski,PAULING,Niu}. 
Thus, we include the binaries Mo$_2$C, Fe$_2$Mo and Fe$_3$C \cite{Massalski,PAULING}. 
In addition, since our Fe-rich Mo:Fe experiments were performed with compositions close to 
Fe$_4$Mo \cite{Avetik_APL_2007}, we include a random phase Fe$_4$Mo generated with the special
quasi-random structure formalism (SQS).
Bulk ternary carbides, which have been widely investigated due to their importance in alloys and steel,
can be considered as derivatives of binary structures with extra C atoms in the interstices
of the basic metal alloy structures. Three possible ternary phases have been
reported for bulk Fe-Mo-C \cite{Rivlin} and they 
are referred as $\tau_1$ ($M$$_6$C), $\tau_2$ ($M$$_3$C) and $\tau_3$ ($M$$_{23}$C$_6$) ($M$ is the metal species). 
For simplicity, we follow the same nomenclature.  $\tau_1$ is the
well-known $M$$_6$C phase, which has been observed experimentally
as Fe$_4$Mo$_2$C and Fe$_3$Mo$_3$C structures ($\eta$ carbides) \cite{RivlinRef31,RivlinRef32}. 
Both of these structures are fcc \cite{Rivlin} but have
different lattice spacings \cite{Pearson}.  
Our calculations show that the most stable variant $\tau_1$ is 
Fe$_4$Mo$_2$C, and we denote
it as $\tau_1$ henceforth.  $\tau_2$ is the Fe$_2$MoC phase, which has
an orthorhombic symmetry distinct from that of Fe$_3$C
\cite{RivlinRef33,Kuo}. 
We consider Fe$_{21}$Mo$_2$C$_6$ \cite{Krainer} as the third $\tau_3$ fcc phase.
We use the Cr$_{23}$C$_6$ as the prototype structure \cite{Pearson}
where Fe and Mo substitute for Cr. Although $M$$_{23}$C$_6$ type phases do
not appear in the stable C-Fe or C-Mo systems, they have been reported 
in ternary C-Fe-Mo systems and also appear as transitional
products in solid state reactions \cite{Rivlin}.
Time-temperature precipitation diagrams of low-C steels have identified 
$\tau_2$-$M$$_3$C, $\tau_3$-$M$$_{23}$C$_6$ and $\tau_1$-$M$$_6$C as
low-temperature, metastable and stable carbides, respectively \cite{Janovec}.
Furthermore, $\tau_2$-$M$$_3$C carbides precipitate quickly due to carbon-diffusion 
controlled reaction while $\tau_3$-$M$$_{23}$C$_6$ carbides precipitate due to
substitutional-diffusion controlled reactions. 
The latter phenomenon, requiring high temperature, longer times and producing metastable phases \cite{Janovec} 
is not expected to enhance the catalytic deactivation of the nanoparticle.
In summary, as long as the presence of carbon does not lead to excessive
formation of $M$$_3$C (Fe$_3$C and  $\tau_2$-Fe$_2$MoC), the catalyst should remain catalytically active for SWCNT growth.

{\bf Results.}
A structure at a given composition is considered stable (at zero temperature 
and without zero-point motion) if it has the lowest formation energy of all 
structures at this composition and, if on the ternary phase diagram, it lies below 
the {\it convex hull of tie planes} connecting all the other stable structures \cite{SC13,SC20,SC21,SC26}.
Phases lying above the convex hull and with {\it small} positive formation energies may be favored at higher temperatures due to 
configurational and vibrational entropy contributions.
\begin{figure}[tb]
  \begin{center}
    \centerline{\epsfig{file=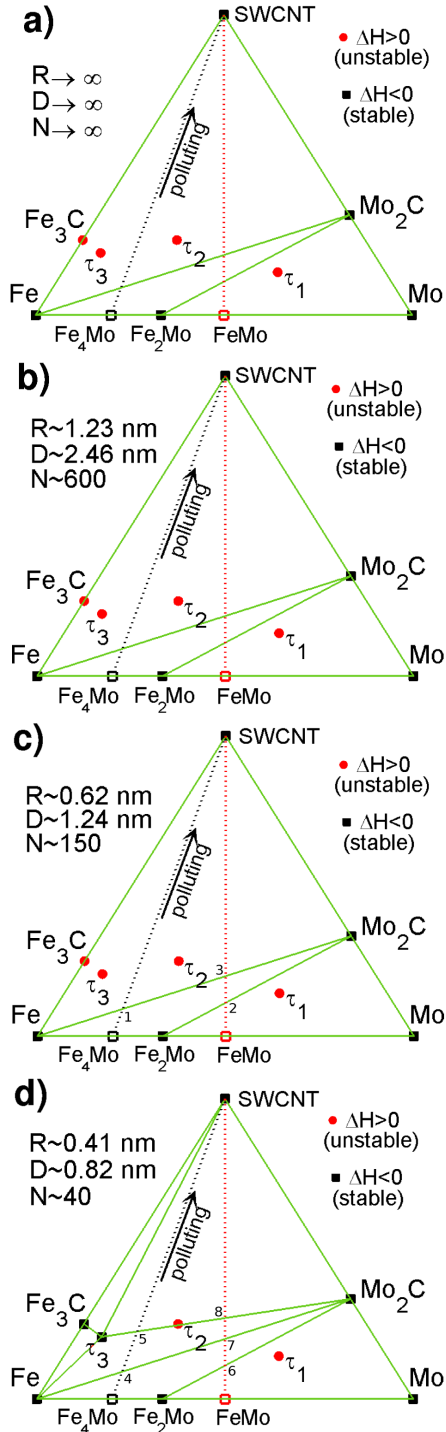,height=192mm,clip=}}
    \vspace{-5mm}
    \caption{\small
       (color online). 
     Ternary phase diagram for Fe-Mo-C nanoparticles of $R\sim\infty,1.23,0.62,0.41$ nm. 
      Notations are explained in the text. }
     \label{figure1}
  \end{center}
  \vspace{-13mm}
\end{figure}

We generate the convex hulls with the {\sf qconvex} package \cite{QHull}.
By projecting the 3D facets onto the 2D plane we obtain the graphs shown 
in Figure \ref{figure1}, where the panels (a),(b),(c) and (d) represent the phase diagrams at zero temperature
of nanoparticles of radii $R\sim\infty,1.23,0.62,0.41$ nm, calculated 
at $P=0,2,4,$ and 6 GPa, respectively.
Stable and unstable phases are shown as black squares and red dots, respectively.
The solid green lines, connecting the stable phases are the projected edges of the 3D convex hull ``facets''.
The numbers 1$\dots$8 in panels (c) and (d) indicate the boundary phases' crossings of our test cases, and are discussed below.
Table \ref{table1} shows the formation energies of the various competing
phases calculated with respect decomposition into the appropriate 
stable components (reported in the ``refs.'' lines).

{\footnotesize
  \begin{table}[htq]
    \centering
    \begin{tabular}{c | c | c | c | c}
      \hline\hline
      R (nm)  & $\infty$ & 1.23 & 0.62 & 0.41\\
      P (GPa) & 0 & 2 & 4 & 6 \\
      N (\#)  & $\infty$ & $\sim$600& $\sim$150 & $\sim$40\\ \hline\hline
      Phases & (meV/at.) & (meV/at.) & (meV/at.) & (meV/at.) \\
      \hline
      Fe$_3$C   & 48.7 & 28.9 & 6.0  & -20.9\\
      (refs.)    & {\footnotesize Fe,C} & {\footnotesize Fe,C} & {\footnotesize Fe,C} & {\footnotesize Fe,C} \\
      \hline
      Mo$_2$C & -113.7& -131.3& -154.3& -185.9\\
      (refs.)    & {\footnotesize Mo,C} & {\footnotesize Mo,C} & {\footnotesize Mo,C} & {\footnotesize Mo,C} \\
      \hline
      Fe$_2$Mo & -2.0 & -6.9& -14.1& -21.2\\
      (refs.) & {\footnotesize Fe,Mo} &  {\footnotesize Fe,Mo} &  {\footnotesize Fe,Mo} &  {\footnotesize Fe,Mo} \\
      \hline
      $\tau_1$& 46.8& 47.2 & 47.2 & 47.2 \\
      (refs.) & {\footnotesize Fe$_2$Mo,} & {\footnotesize Fe$_2$Mo,} & {\footnotesize Fe$_2$Mo,} & {\footnotesize Fe$_2$Mo,} \\
      (refs.) & {\footnotesize Mo$_2$C,Mo} & {\footnotesize Mo$_2$C,Mo} & {\footnotesize Mo$_2$C,Mo} & {\footnotesize Mo$_2$C,Mo} \\
      \hline
      $\tau_2$ & 451.7 &  433.4 & 413.2 & 414.6  \\
      (refs.) & {\footnotesize Fe,C,} & {\footnotesize Fe,C,} & {\footnotesize Fe,C,} & {\footnotesize $\tau_3$,C,} \\
      (refs.) & {\footnotesize Mo$_2$C} & {\footnotesize Mo$_2$C} & {\footnotesize Mo$_2$C} & {\footnotesize Mo$_2$C} \\
      \hline
      $\tau_3$ & 61.4 & 12.4 & 7.6 & -14.4 \\
      (refs.) & {\footnotesize Fe,C,} & {\footnotesize Fe,C,} & {\footnotesize Fe,C,} & {\footnotesize Fe,Fe$_3$C,} \\
      (refs.) & {\footnotesize Mo$_2$C} & {\footnotesize Mo$_2$C} & {\footnotesize Mo$_2$C} & {\footnotesize Mo$_2$C} \\
      \hline\hline
    \end{tabular}
    \vspace{-2mm}
    \caption{\small Formation energies (meV/atom) for binary and ternary
      phases for nanocatalysts of different sizes: Fe$_3$C, Mo$_2$C,
      Fe$_2$Mo, $\tau_1$, $\tau_2$ and $\tau_3$, calculated with respect
      to the reference species shown in the table.}
    \label{table1} 
  \end{table}
}

In each panel of Figure \ref{figure1},
the two dotted lines connecting Fe$_4$Mo and FeMo to SWCNT 
denote the introduction of carbon into the system.
Fe$_4$Mo has been reported to be an effective catalyst composition \cite{Avetik_APL_2007}
while FeMo represents a hypothetical Fe:Mo particle with a Mo content larger than 33\%.  This Mo-rich nanoparticle
gives low yields of SWCNTs at high temperature \cite{note_Avetik}.
In thermodynamic terms, the growth of SWCNTs has to be considered as nucleation of ordered C, hence
the balance between all the competing phases along the dotted lines controls the activity of the particle.
The size of the particles have been calculated with the interpolation shown in 
Figure 2 of Ref. \cite{Curtarolo_SOLUBILITY1,Curtarolo_SOLUBILITY2}. 
We investigate a very small particle with $N_{atoms}\sim40$, which may 
able to support SWCNT growth \cite{kimX}.
 to explore size-induced stabilization trends.
As seen from the zero temperature phase diagrams in Figure \ref{figure1} 
and the energies in Table \ref{table1},
Mo$_2$C and Fe$_2$Mo are stable for bulk materials as well as nanoparticles,
while Fe$_3$C and $\tau_3$ are stable only for small nanoparticles of radius $R\sim0.41$ nm
(E$_f$[Fe$_3$C]$\le0$ for $R<0.58$ nm \cite{Curtarolo_SOLUBILITY1,Curtarolo_SOLUBILITY2}).
$\tau_1$ and $\tau_2$ are always unstable.

\begin{figure}[htb]
  \begin{center}
    \centerline{\epsfig{file=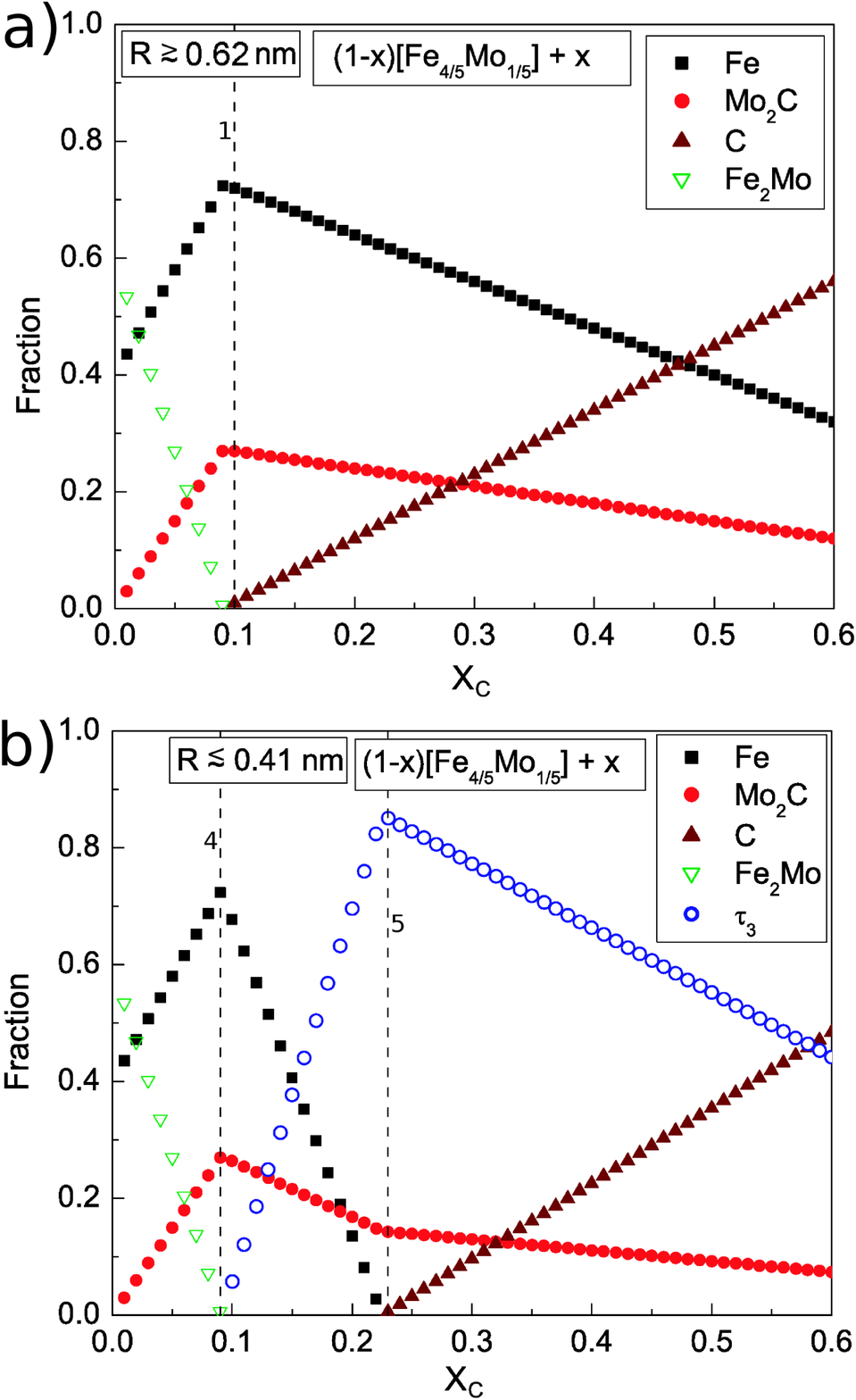,width=78mm,clip=}}
    \vspace{-4mm}
    \caption{\small
      (color online). 
      Panel {\bf a)}.
      Fraction of species for catalyst composition 
      [Fe$_{4/5}$Mo$_{1/5}$]$_{1-x}$-C$_x$ and nanocatalyst of $R\gtrsim$ 0.62 nm. 
      The dashed vertical line, labeled as ``1'', represents [Fe$_{4/5}$Mo$_{1/5}$]$_{1-x}$-C$_x$
      crossing the boundary phase Fe$\leftrightarrow$Mo$_2$C, as shown in Figure \ref{figure1}(c). 
      Panel {\bf b)}.
      Fraction of species for catalyst composition 
      [Fe$_{4/5}$Mo$_{1/5}$]$_{1-x}$-C$_x$ and nanocatalyst of $R\lesssim$ 0.41 nm.
      The dashed vertical lines, labeled as ``4'' and ``5'', represent 
      [Fe$_{4/5}$Mo$_{1/5}$]$_{1-x}$-C$_x$
      crossing the boundary phases Fe$\leftrightarrow$Mo$_2$C and $\tau_3$$\leftrightarrow$Mo$_2$C, 
      as shown in Figure \ref{figure1}(d). }
    \label{figure2}
  \vspace{-6mm}
   \end{center}
\end{figure}

{\it Fe$_4$Mo particles}.
The most obvious advantage that a Fe$_4$Mo particle has over a pure Fe particle 
is that the [Fe$_{4/5}$Mo$_{1/5}$]$_{1-x}$-C$_x$ line does not intercept any
carbide (Fe$_3$C, $\tau_3$,$\tau_2$). This implies that, at least at low temperatures, there 
is a surplus of unbounded metal (probably even at high temperatures since the line is far from all of the competing stable phases).
This is illustrated in Figure \ref{figure2}, which shows the fractional evolution of species as one progresses along the 
[Fe$_{4/5}$Mo$_{1/5}$]$_{1-x}$-C$_x$ line in Figure \ref{figure1}.  

{\it Large Fe$_4$Mo particles ($R\gtrsim$ 0.62 nm)}.
When a large Fe$_4$Mo particle ($R\gtrsim$ 0.62 nm in Figure \ref{figure2} (a)) is exposed to carbon feedstock,
the Mo$_2$C phase nucleates by dissociating Fe$_2$Mo. This is seen for concentrations 
between $0<x_c\lesssim0.09$. The vertical dashed line ``1'' in the figure indicates the 
boundary crossing of the [Fe$_{4/5}$Mo$_{1/5}$]$_{1-x}$-C$_x$ pollution path
with the Fe$\leftrightarrow$Mo$_2$C edge, as shown in Figure \ref{figure1}(c).
Nucleation of Mo$_2$C has two consequences: 
it consumes carbon atoms that are added to the particle by bonding them to molybdenum, 
and it releases free-Fe which is beneficial 
for the catalytic activity of the cluster (the Fe fraction increases with C concentration in the range $0<x_c\lesssim0.09$).
In addition, while forming Mo$_2$C, carbon does not produce SWCNTs. 
In fact, we need to saturate the particle in C
and enter the region Fe-Mo$_2$C-SWCNT of the phase diagram in Figure \ref{figure1}(c) 
before free carbon is available.  Hence, the fraction of C is non-zero only after $x_c\sim0.09$ in
Figure \ref{figure2}(a). For $x_c\lesssim0.09$ the free Fe is expected 
to be on the particle surface since the fraction of free-Fe is larger than 
that of Mo$_2$C and, as presented later, 
surface energy calculations show that free-Fe resides  at the surface of the particle whereas the 
Mo (Mo$_2$C) parties found near particle core. 
To conclude, steady state growth of SWCNTs is possible from large Fe$_4$Mo particles since free, catalytically active  
Fe is present, even for $x_c\gtrsim0.09$.

{\it Small Fe$_4$Mo particles ($R\lesssim$ 0.41 nm)}.
Similarly to the large Fe$_4$Mo particle, initial exposure of  
the small Fe$_4$Mo cluster ($R\lesssim$ 0.41 nm) to carbon feedstock promotes Mo$_2$C 
nucleation by dissociating Fe$_2$Mo and releasing free Fe.
However, for $x_c\gtrsim 0.09$ the scenario is different.
In fact, while the fraction of free Fe increases between $0<x_c\lesssim0.09$ (left of line ``4'' in Figure \ref{figure2} (b)), it 
reduces to zero between $0.09\lesssim x_c \lesssim 0.23$ (points ``4'' and ``5'' of  Figure \ref{figure1}(d)).
Concomitantly, the carbide $\tau_3$ nucleates by bonding carbon atoms that are added to the 
particle and hence eliminating the amount of free Fe that was released during the growth of Mo$_2$C.
Similarly to Fe-C particles \cite{Curtarolo_SOLUBILITY1,Curtarolo_SOLUBILITY2} with the carbide Fe$_3$C,
the nucleation of $\tau_3$ is clearly detrimental for SWCNT growth.
In fact, the lack of simultaneous presence of free-Fe and excess C 
causes the particle to be catalytically inactive. 
However, the different formation energies of $\tau_3$ and Fe$_3$ lead to different size 
thresholds between the two regimes (growth/no-growth).
In Refs. \cite{Curtarolo_SOLUBILITY1,Curtarolo_SOLUBILITY2} we estimated that the minimum 
radius needed for a low-temperature Fe particle to be active is $R_{min}^{Fe}\sim 0.58$ nm.
For the Fe$_4$Mo case, 
by interpolating the energies of $\tau_3$ reported in Table \ref{table1},
by determining $P$ at which $E[\tau_3(P)]=0$, 
and by using the size/pressure approximation relation (\ref{sizepressureequation}), 
we obtain $R_{min}^{Fe_4Mo}\sim0.52$ nm. 
The reduced size advantage $R_{min}^{Fe_4Mo}<R_{min}^{Fe}$ agrees with experimental results that show that nanotubes, 
grown from Fe-rich Fe:Mo catalysts are thinner than those grown from pure Fe particles \cite{Avetik_APL_2007}.
As pointed out before \cite{Curtarolo_SOLUBILITY1,Curtarolo_SOLUBILITY2}, the presence of $R_{min}^{Fe_4Mo}$ defines three 
possible thermodynamic scenarios: steady-state-, limited- and no-growth for
$R > R_{min}^{Fe_4Mo}$, $R\sim R_{min}^{Fe_4Mo}$ and $R< R_{min}^{Fe_4Mo}$, respectively.
It is worth mentioning that as long as the concentration of Mo in a Fe-rich Fe:Mo clusters is chosen to be 
slightly larger than that of $\tau_3$ we obtain $R_{min}^{Fe:Mo}<R_{min}^{Fe}$. 

{\it FeMo test particles}.
To address the poor growth capability of Mo-rich Fe:Mo catalysts 
(low yield of SWCNTs at high temperature \cite{note_Avetik})
we repeated the above analysis with an equiconcentration 
FeMo particle. 
For simplicity, we choose to avoid dealing with the Mo$_{5.1}$Feo$_{7.9}$ 
phase reported at low temperature \cite{Massalski,PAULING} because, 
as our results show, its presence does not affect the 
pollution mechanism discussion.

\begin{figure}[htp]
  \begin{center}
    \centerline{\epsfig{file=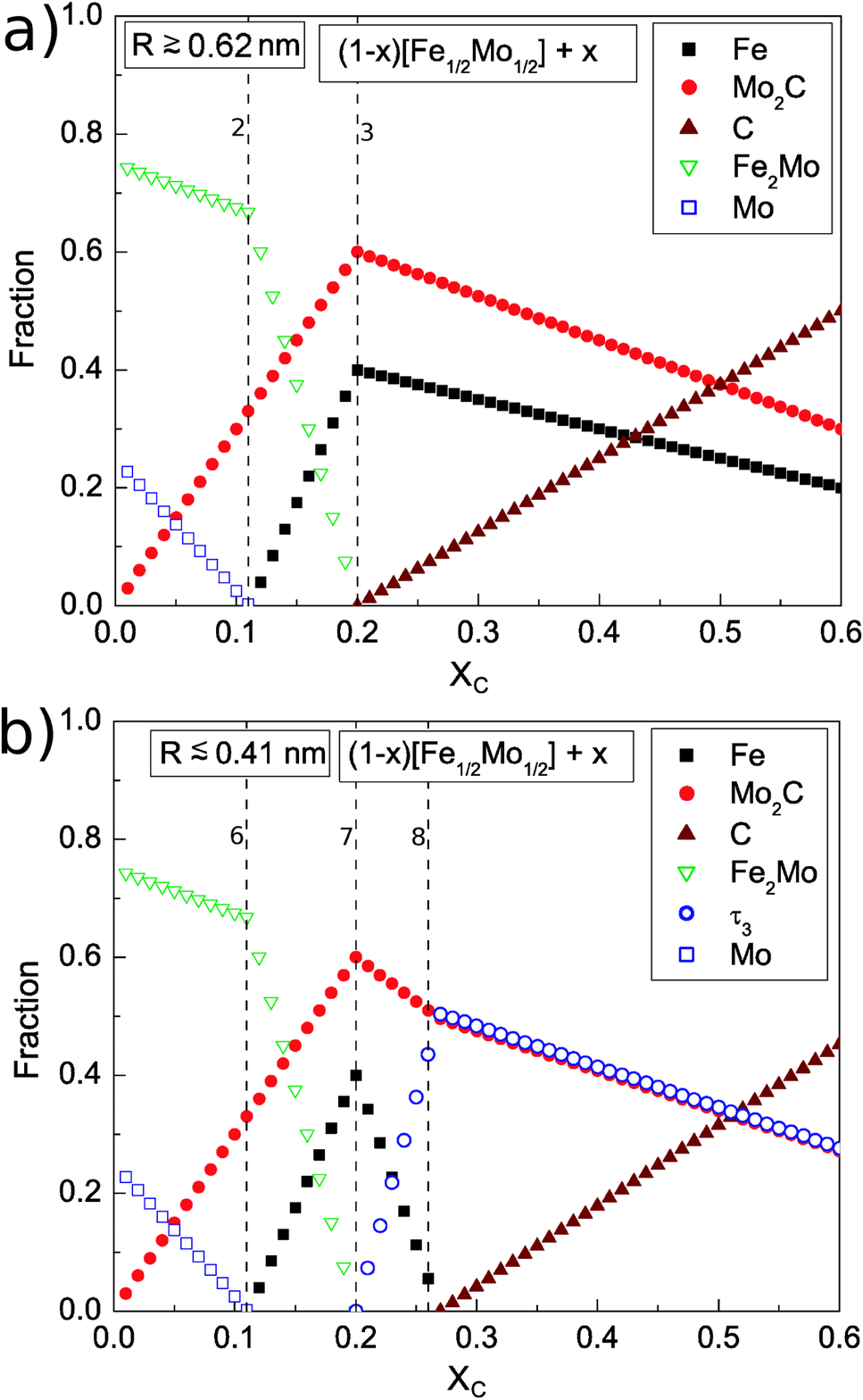,width=80mm,clip=}}
    \caption{\small 
      (color online). 
      Panel {\bf a)}.
      Fraction of species for catalyst composition 
      [Fe$_{1/2}$Mo$_{1/2}$]$_{1-x}$-C$_x$ and nanocatalyst of R $\gtrsim$ 0.62 nm. 
      The dashed vertical lines, labeled as ``2'' and ``3'', represent 
      [Fe$_{1/2}$Mo$_{1/2}$]$_{1-x}$-C$_x$
      crossing the boundary phases Fe$_2$Mo$\leftrightarrow$Mo$_2$C and Fe$\leftrightarrow$Mo$_2$C, 
      as shown in Figure \ref{figure1}(c). 
      Panel {\bf b)}.
      Fraction of species for catalyst composition
      [Fe$_{1/2}$Mo$_{1/2}$]$_{1-x}$-C$_x$ and nanocatalyst of R $\lesssim$ 0.41 nm. 
      The dashed vertical lines, labeled as ``6'', ``7'' and ``8'', represent 
      [Fe$_{1/2}$Mo$_{1/2}$]$_{1-x}$-C$_x$
      crossing the boundary phases 
      Fe$_2$Mo$\leftrightarrow$Mo$_2$C, Fe$\leftrightarrow$Mo$_2$C, and $\tau_3$$\leftrightarrow$Mo$_2$C,
      as shown in Figure \ref{figure1}(d). }
    \label{figure3}
  \end{center}
  \vspace{-4mm}
\end{figure}

{\it Large FeMo particles ($R\gtrsim$ 0.62 nm)}.
Contrary to Fe$_4$Mo particles, large FeMo clusters contain ample amounts of molybdenum 
capable of nucleating Mo$_2$C. Irrespective of whether Mo$_{5.1}$Feo$_{7.9}$ is included in the discussion,
both free-Fe and an excess of C are present
for $x_c\gtrsim 0.20$ (line ``3'' in Figure \ref{figure3}(a) and phase boundary ``3'' in Figure \ref{figure1}(c)).
Hence, the growth SWCNTs is possible for $x_c\gtrsim 0.20$.
However, since the fraction of Fe considerably smaller than that of Mo$_2$C,
the activity of the whole FeMo particle is drastically smaller than that of the Fe$_4$Mo catalyst.
Thus, although it is possible to growth SWCNTs from large FeMo clusters, the expected yield is 
low and the synthesis temperature needs to be high (to overcome the reduced fraction of catalytically active free-Fe),
as experimentally reported \cite{note_Avetik}.
Thermodynamically, an excess of C and a amount of free-Fe for $x_c\gtrsim 0.20$ guarantees 
the existence of steady-state growth of SWCNTs (albeit slow and inefficient).

{\it Small FeMo particles ($R\lesssim$ 0.41 nm)}.
Small FeMo particles are similar to small Fe$_4$Mo clusters.
Nucleation of $\tau_3$ and the absence of free-Fe and excess C means 
that the particle is catalytically inactive. 
The minimum cluster radius, $R_{min}^{FeMo}$ is the same as $R_{min}^{Fe_4Mo}$ because 
this quantity is determined by the stabilization of the same $\tau_3$ phase. 
Thus, there are three possible thermodynamic scenarios that are similar to the previous case, i.e., 
low-yield steady-state-, 
low-yield limited- and no-growth for
$R > R_{min}^{FeMo}$, $R\sim R_{min}^{FeMo}$ and $R< R_{min}^{FeMo}$, respectively.
The analysis can be extended to different fractions of Fe and Mo and the results are summarized in Table \ref{table2}.
{\footnotesize
  \begin{table}[htq]
    \centering
    \begin{tabular}{c | c | c }
      \hline\hline
      catalyst   & Fe-rich Fe$_{1-x}$Mo$_x$ &  Mo-rich Fe$_{1-x}$Mo$_x$ \\
      $x_c$ range     & $x^\star_{\tau_3}<x<1/3$ & $x>1/3$       \\        
      (test)    & Fe$_4$Mo & FeMo      \\ \hline
      $R>R_{min}^{FeMo}$     & {\footnotesize steady-state growth} & {\footnotesize low-yield steady-state growth} \\
      $R\sim R_{min}^{FeMo}$ & {\footnotesize limited growth} & {\footnotesize low-yield limited growth}  \\
      $R<R_{min}^{FeMo}$     & {\footnotesize  no growth} & {\footnotesize  no growth} \\
      \hline\hline
    \end{tabular}
     \caption{\small 
      Scenarios of thermodynamically allowed
      SWCNT growth modes 
      for different sizes and compositions of Fe:Mo catalysts.
      $x^\star_{\tau_3}$ is the concentration of Mo in $\tau_3$ removed of all C.
      $R_{min}^{FeMo}=0.52 <R_{min}^{Fe}=0.58$ nm from Refs.
      \cite{Curtarolo_SOLUBILITY1,Curtarolo_SOLUBILITY2}. 
    }
    \label{table2} 
  \end{table}
}

{\bf Surface energies.}
The chemical species that are found on the surface of a Fe:Mo cluster with coexisting 
Fe, 
and Fe$_2$Mo phases
is the species that has the lowest 
surface energy per unit area, $\gamma$.
This quantity can be obtained from the relation $\gamma=\left[E_n-nE_B\right]/{2 A}$,
where $E_n$ is the total energy of a $n$-layer slab, $E_B$ is the total
energy of a single bulk layer, and $A$ is the surface of the unit cell 
(the factor two accounts for the creation of two surfaces) \cite{Blonski}. 
For bcc Fe, we construct slabs with the lowest surface energy (1 1 0) termination \cite{Blonski}. 
Surface termination for the ordered Fe$_2$Mo phase has not
been reported in experiments or in theoretical works. 
By using the package {\small AFLOW}, which performs high-throughput simultaneous 
optimization of planar density (high) and number of broken bonds (low) \cite{Blonski,AFLOW},
we find that the closed packed (0 0 4) plane of Fe$_2$Mo has the lowest energy.
We obtain $\gamma_{Fe}^{(1 1 0)}=2.44$ J/m$^2$ and $\gamma_{Fe_2Mo}^{(0 0 4)}=3.12$ J/m$^2$.
The relation $\gamma_{Fe} < \gamma_{Fe_2Mo}$ indicates that Mo
will not be at the cluster surface.
Thus, in Fe-rich Fe:Mo nanocatalysts, Fe covers as much surface area as possible, 
and the aforementioned nucleation of Mo$_2$C caused by C pollution is advantageous 
for the activity of the particle, by releasing free-Fe on the surface.


{\bf Conclusions.} In this paper we discuss
the role of Mo in the thermodynamic properties 
of Fe:Mo nanocatalysists by using the size-pressure approximation 
and {\it ab initio} modeling.
We show that for both Fe-rich ($\sim$ 80\% Fe or more) 
and Mo-rich ($\sim$ 50\% Mo or more) Fe:Mo clusters, 
the presence of carbon causes nucleation of Mo$_2$C which 
enhances the activity of the particle by releasing free-Fe.
With respect to pure Fe-catalysts, the addition of Mo (up to small 
concentrations) decreases the size of the smallest catalyst needed for low-temperature CVD 
steady-state growth to $R_{min}^{Fe:Mo}\sim0.52$ nm. 

{\bf Acknowledgement.} We acknowledge helpful discussions with 
A. Kolmogorov, N. Li, T. Tan, E. Mora and F. Cervantes-Sodi.
The authors are grateful for computer time allocated 
at the Teragrid facilities. 
This research was supported by Honda Research Institute USA, Inc. 
SC is supported by ONR (N00014-07-1-0878) and NSF (DMR-0639822).

%


\begin{thebibliography}{99}
\bibitem{noteFeMo1}
  Our calculations for bulk bc- Fe show that Mo prefers a substitutional position over
  interstitial. The calculation was done for a bcc Fe unitcell (3x3x3)
  with 27 atoms in which one Fe was substituted by Mo. 
  Formation energies are calculated with the usual framework \cite{SC13,SC26}.
\end{thebibliography}
\end{document}